\documentclass[journal]{IEEEtran}
\usepackage{amsmath,amsfonts}
\usepackage{algorithmic}
\usepackage{algorithm}
\usepackage{array}
\usepackage{bm}
\usepackage{xcolor}
\usepackage[caption=false,font=normalsize,labelfont=sf,textfont=sf]{subfig}
\usepackage{textcomp}
\usepackage{stfloats}
\usepackage{url}
\usepackage{booktabs}
\usepackage{verbatim}
\usepackage{graphicx}
\usepackage{cite}
\usepackage{xspace}
\usepackage[hidelinks]{hyperref}
\usepackage{orcidlink}
\newcommand{\Ours}{Arctic E2E\xspace}
\begin{document}

\title{\Ours: An End-to-End Framework for Forecasting Arctic Sea Ice and Upper-Ocean States from Heterogeneous Observations}

\author{Mingqiang Wang\,\orcidlink{0009-0006-2551-5758},  Xiubao Yi\,\orcidlink{0009-0002-6693-0078}, Yanhai Gan\,\orcidlink{0000-0002-5315-2740}, Feng Gao\,\orcidlink{0000-0002-1825-328X} and Junyu Dong\orcidlink{0000-0001-7012-2087}

\thanks{This work was supported in part by the Key R\&D Program of Shandong Province under Grant 2025CXPT185, in part by the Natural Science Foundation of China under Grant 42406192, in part by the Fundamental Research Funds for the Central Universities under Grant 202413040, and in part by the Postdoctoral Project of Qingdao under Grant QDBSH20240102021.}

\thanks{M. Wang, X.Yi, Y. Gan, F. Gao and J. Dong are with the State Key Laboratory of Physical Oceanography and also with the School of Computer Science and Technology, Ocean University of China, Qingdao, 266100, China(e-mail: wangmingqiang@stu.ouc.edu.cn, ganyanhai@ouc.edu.cn, dongjunyu@ouc.edu.cn)}

\thanks{Corresponding authors: Yanhai Gan and Junyu Dong.}

}

\markboth{Submmited to IEEE}%
{Wang \MakeLowercase{\textit{et al.}}: Arctic E2E: An End-to-End Framework for Forecasting Arctic Sea
Ice and Upper-Ocean States from Heterogeneous Observations}


\maketitle

\begin{abstract}
Accurate short-term forecasts of Arctic sea ice and upper-ocean conditions are essential for polar navigation and offshore operations. Most data-driven forecasting models are initialized from gridded reanalysis or retrieval products, which constrains initialization timeliness and decouples observation processing from forecast objectives. We propose \Ours, an end-to-end framework that forecasts Arctic sea-ice and upper-ocean states directly from heterogeneous observations. From seven days of gridded AMSR and SSMIS brightness temperatures and sparse in situ temperature--salinity profiles, a density-aware observation encoder reconstructs historical physical states, a Swin Transformer-based processor predicts the following seven days in a single pass, and two parallel decoders perform multiscale grid refinement and coordinate-conditioned point prediction. The model jointly forecasts sea-ice concentration (SIC), sea-ice thickness (SIT), and potential temperature and salinity at ten levels in the upper 0--40~m on a 12-km grid, and supports coordinate-conditioned temperature and salinity queries within the modeled upper ocean. Staged pretraining followed by joint optimization allows forecast errors to update the observation encoder. Training uses GLORYS12 reanalysis for processor pretraining, gridded supervision, and profile augmentation, whereas inference initializes forecasts solely from historical satellite brightness temperatures and available in situ profiles without a gridded reanalysis initial state. Across 961 validation initializations from 2023 to 2025, \Ours achieves the lowest errors in six of eight metrics against six observation-to-forecast baselines, reducing seven-day window-pooled SIC and SIT root-mean-square error (RMSE) by 32.0\% and 22.3\%, respectively, relative to the best baseline for each metric. The point decoder reduces salinity RMSE by 23.2\% at matched profile samples relative to trilinear interpolation of refined gridded forecasts, and day-7 full-domain SIC RMSE is lower than observation-encoder persistence in all four seasons.
\end{abstract}

\begin{IEEEkeywords}
Arctic sea ice forecasting, end-to-end learning, satellite brightness temperature, in situ profiles, upper-ocean temperature and salinity, deep learning
\end{IEEEkeywords}

\section{Introduction}

Accurate forecasts of Arctic sea ice are essential for polar route planning, scientific expeditions, and offshore operations. In recent years, Arctic sea ice has undergone substantial changes in thickness and age structure, while summer ice cover continues to decline~\cite{sumata2023regime,heuze2024icefree}. Although retreating ice alters operational conditions, multiyear ice transport into shipping lanes and wind-driven drift and convergence can still obstruct navigation and trap vessels~\cite{cook2024chokepoints,moore2026entrapment}. Timely forecasts of SIC and SIT over the coming days are therefore critical, and have been pursued through numerical modeling, statistical methods, and deep learning.

Coupled sea-ice--ocean and atmosphere--sea-ice--ocean numerical models simulate ice dynamics and thermodynamics explicitly, relying on data assimilation for initialization; for example, Liu et al.\ assimilated SIC and SIT into FGOALS-f2 to improve melt-season predictions~\cite{liu2025summerthickness}. However, numerical forecasts remain sensitive to initial conditions and parameterizations while incurring high computational costs at fine resolutions. Statistical methods offer a lightweight alternative by fitting empirical relationships to historical records---such as SpectraCast~\cite{spectracast2026}, which decomposes periodic SIC components and autoregresses residuals---yet struggle with complex spatial interactions under a shifting climate state. Deep learning flexibly captures nonlinear spatiotemporal patterns from historical sea-ice and environmental fields across timescales, including twelve-month U-Net SIC predictions~\cite{kim2025longterm}, one-to-three-day high-resolution regional forecasts fusing ice charts and satellite products~\cite{kvanum2025shortterm}, and post-processing corrections to TOPAZ4 forecasts~\cite{palerme2024correction}.

Despite these advances, existing data-driven forecasting pipelines face three key limitations:
\emph{1)~Decoupled observation processing and reanalysis-dependent initialization:} Most predictors initialize from pre-generated satellite retrieval or reanalysis products, which carry inherent estimation uncertainties~\cite{wernecke2024uncertainty} and production delays. Because upstream retrieval or analysis is isolated from the forecasting network, downstream forecast errors cannot optimize how features are extracted from observational inputs.
\emph{2)~Modality gap between surface satellites and subsurface profiles:} Sea-ice evolution is governed not only by surface conditions but also by upper-ocean heat transport and temperature--salinity (T/S) stratification~\cite{polyakov2023atlantification,xu2024bering,polyakov2025atlantification}, while ice retreat reciprocally modulates upper-ocean heating~\cite{barkhordarian2024heatwaves}. Satellite microwave sensors offer daily pan-Arctic coverage without subsurface visibility, whereas sparse in situ profiles capture vertical ocean structure~\cite{anhaus2025underice} at irregular locations and depths, complicating their direct joint ingestion.
\emph{3)~Discrepancy between gridded states and local in situ observations:} Gridded reanalyses smooth sharp spatial transitions and local water-column variations, so interpolating gridded forecasts to observation locations can degrade local accuracy. While recent end-to-end weather models such as Aardvark Weather~\cite{aardvark2025} and FuXi Weather~\cite{fuxi2025} demonstrate that observation encoding and forecasting can be trained jointly for atmospheric states, an end-to-end observation-to-forecast framework combining satellite brightness temperatures and sparse subsurface profiles for coupled Arctic sea-ice and upper-ocean prediction remains underexplored.

To address these challenges, we propose \Ours, an end-to-end framework that forecasts seven-day Arctic sea-ice and upper-ocean (0--40~m) states from seven days of gridded AMSR and SSMIS microwave brightness temperatures and sparse in situ T/S profiles. Training uses GLORYS12 reanalysis for processor pretraining, gridded supervision, and profile augmentation; at inference, \Ours initializes forecasts solely from historical observations without gridded reanalysis initial conditions or external weather forcing. Our main contributions are as follows:
\begin{itemize}
    \item \textbf{End-to-End Heterogeneous Observation-to-Forecast Pipeline:} We integrate density-aware SetConv observation encoding, Swin-based direct multi-step temporal prediction, and dual-branch decoding within a jointly trainable network. Staged pretraining followed by end-to-end fine-tuning allows historical reconstruction, future gridded, and pointwise objectives to jointly optimize upstream observation representations.
    \item \textbf{Surface--Subsurface Observation Fusion and Parallel Dual Decoding:} The encoder fuses track-gapped satellite brightness temperatures with irregular T/S profiles into 22-channel historical physical states. Two parallel decoder heads separate Sobel edge-guided multiscale spatial grid refinement (for 12-km SIC, SIT, and 10-level T/S fields) from coordinate-conditioned T/S queries, reducing SIC and SIT spatial-gradient errors after grid refinement and improving local salinity accuracy at matched profile samples.
    \item \textbf{Comprehensive Multivariate and Pointwise Evaluation:} Across 961 validation initializations (2023--2025) against six observation-to-forecast baselines, \Ours ranks first in six of eight metrics, reducing window-pooled SIC and SIT RMSE by 32.0\% and 22.3\% relative to the best baseline for each metric, reducing pointwise salinity RMSE by 23.2\% over refined-grid trilinear interpolation, and achieving lower day-7 full-domain SIC RMSE than observation-encoder persistence in all four seasons.
\end{itemize}

\section{Related Works}

\noindent\textbf{Sea Ice Forecasting Based on Reanalysis Data.} Gridded reanalysis and retrieval products underpin most deep learning sea-ice models. FCNet incorporates frequency-domain filtering with convolutional layers to preserve high-frequency SIC variations~\cite{fcnet2025}, while SIFusion shares a Swin Transformer encoder--decoder across daily, weekly, and monthly SIC sequences for multi-timescale prediction~\cite{sifusion2025}. To sharpen coarse-grid ice edges, IceDiff pairs a Swin Transformer predictor with a diffusion downscaling prior~\cite{icediff2025}. TPFMNet integrates neighborhood attention, reparameterized convolutions, and frequency features in a probabilistic U-Net for year-round monthly SIC and uncertainty forecasting~\cite{tpfmnet2026}, and MET-AICE combines AMSR2 SIC with ECMWF atmospheric forecasts in an attention U-Net for ten-day European Arctic operations~\cite{metaice2025}. Unlike predictors initialized from retrieved SIC or reanalysis fields, \Ours initializes inference directly from gridded brightness temperatures and available profiles, coupling observation encoding with forecast optimization.

\noindent\textbf{Multivariate Sea Ice and Ocean Forecasting.} Recent works extend forecasting beyond SIC to multivariate sea-ice and 3D ocean states. CA-PredRNN employs cross-variable attention to jointly forecast SIC and SIT along the Northeast Passage~\cite{capredrnn2026}, GenSIM applies scale-aware Transformers with conditional flow matching to coupled simulation states~\cite{gensim2025}, and Gregory et al.\ predict online sea-ice assimilation increments within the coupled SPEAR model~\cite{spear2026}. For subsurface states, WenHai couples air--sea flux calculations with a Swin Transformer on 3D gridded ocean and atmospheric fields~\cite{wenhai2025}, Neptune combines convolutions with a spherical Fourier neural operator to predict global ocean--ice variables from ORAS5 and ERA5 forcing~\cite{neptune2026}, and 3D-OceanNet reconstructs regional 3D T/S fields from surface temperature and height anomalies with Argo validation~\cite{ocean3dnet2026}. However, these systems require complete 3D analysis states or surface retrieval products as inputs, whereas \Ours jointly ingests satellite brightness temperatures and sparse in situ profiles to forecast surface sea ice and 0--40~m T/S fields.

\noindent\textbf{End-to-End Learning from Observations to Forecasts.} In global weather prediction, Aardvark Weather~\cite{aardvark2025} and FuXi Weather~\cite{fuxi2025} connect multisource observation encoding or learned assimilation directly with atmospheric state forecasting and local decoding. In Arctic sea-ice forecasting, Unicorn feeds microwave brightness temperatures alongside SIC and ice-age products into a neural-ODE U-Net for weekly SIC prediction~\cite{unicorn2025}; IceCastNet initializes a graph Transformer from OSI-SAF, CS2SMOS, and TOPAZ products under ECMWF forcing~\cite{icecastnet2026}; and GraphDOP models spatiotemporal dynamics directly in AMSR2 brightness-temperature observation space~\cite{graphdop2025}. Existing sea-ice approaches thus either use brightness temperatures as auxiliary inputs alongside gridded state products or remain in observation space. In contrast, \Ours maps heterogeneous satellite brightness temperatures and sparse subsurface profiles to seven-day gridded forecasts of SIC, SIT, and 0--40~m T/S as well as coordinate-conditioned T/S predictions at matched profile locations within a single end-to-end network.

\section{Methodology}
\label{sec:method}

\subsection{Overall Architecture}
\label{sec:overview}

\begin{figure*}[t]
    \centering
    \includegraphics[width=0.90\textwidth]{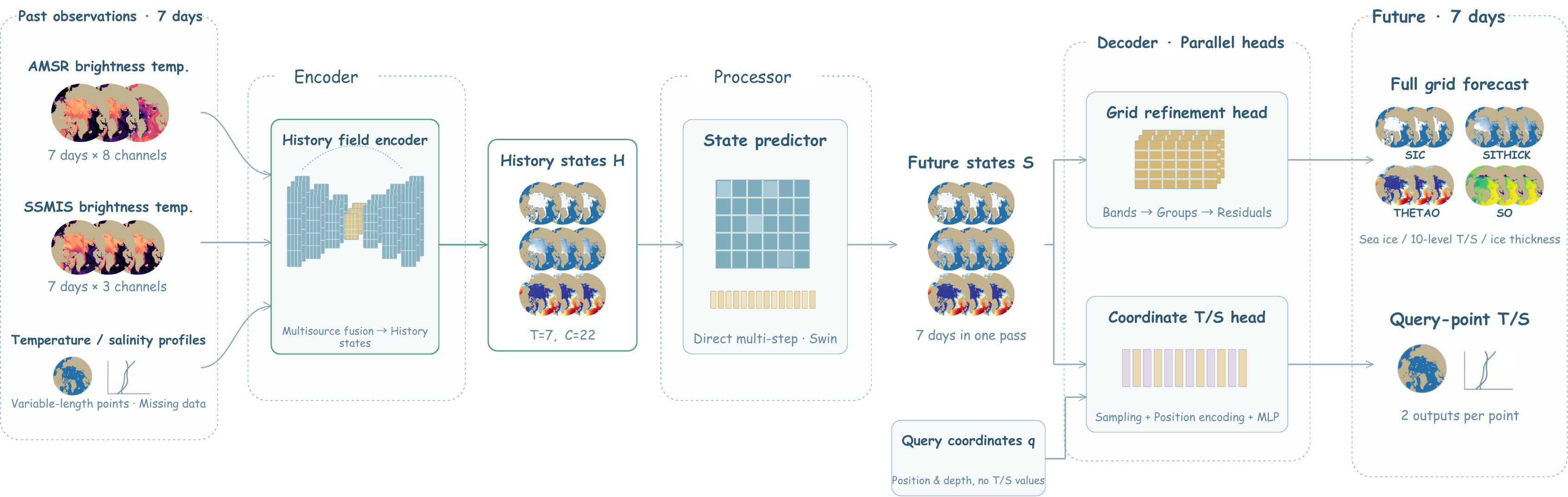}
    \caption{Overview of \Ours. Seven days of AMSR/SSMIS brightness temperatures and irregular T/S profiles are encoded into historical states $\widehat{\mathbf{X}}^{-}$ ($\mathbf{H}$; $T=7, C=22$). The Swin processor predicts all seven future states $\widetilde{\mathbf{X}}^{+}$ ($\mathbf{S}$) in one pass, and parallel decoder heads produce refined gridded forecasts $\widehat{\mathbf{X}}^{+}$ and coordinate-conditioned T/S predictions at queries $\mathcal{Q}^{+}$ ($\mathbf{q}$).}
    \label{fig:overview}
\end{figure*}

\Ours forecasts Arctic sea-ice and upper-ocean states directly from heterogeneous historical observations (Fig.~\ref{fig:overview}). Over the historical window $t\in\{-T_{\mathrm{in}}+1,\ldots,0\}$, available observations are denoted by
\begin{equation}
    \mathcal{O}^{-} = \big( \{\mathbf{A}_t,\mathbf{S}_t, \mathbf{M}^{A}_t,\mathbf{M}^{S}_t\}_{t=-T_{\mathrm{in}}+1}^{0},\, \mathcal{P}^{-} \big),
\end{equation}
where $\mathbf{A}_t\in\mathbb{R}^{8\times H\times W}$ and $\mathbf{S}_t\in\mathbb{R}^{3\times H\times W}$ are AMSR and SSMIS brightness temperatures on the $12$-km Arctic polar stereographic grid ($H=896$, $W=608$) with binary validity masks $\mathbf{M}^{A}_t$ and $\mathbf{M}^{S}_t$, and $\mathcal{P}^{-}$ contains irregularly sampled T/S profiles from ITP and WOD. The target physical state $\mathbf{X}_t\in\mathbb{R}^{C\times H\times W}$ ($C=22$) comprises SIC, SIT, and ocean potential temperature and salinity across ten nominal depth levels $\{0,2,4,6,10,15,20,25,30,40\}$~m, with $T_{\mathrm{in}}=T_{\mathrm{out}}=7$.

The model couples an observation encoder $\mathcal{E}_{\theta}$, a temporal processor $\mathcal{P}_{\phi}$, and parallel gridded ($\mathcal{D}^{g}_{\psi}$) and pointwise ($\mathcal{D}^{p}_{\omega}$) decoder heads:
\begin{equation}
\begin{aligned}
    \widehat{\mathbf{X}}^{-} &= \mathcal{E}_{\theta}(\mathcal{O}^{-}), &
    \widetilde{\mathbf{X}}^{+} &= \mathcal{P}_{\phi}(\widehat{\mathbf{X}}^{-}), \\
    \widehat{\mathbf{X}}^{+} &= \mathcal{D}^{g}_{\psi}(\widetilde{\mathbf{X}}^{+}), &
    \widehat{\mathbf{Y}}^{+} &= \mathcal{D}^{p}_{\omega}(\widetilde{\mathbf{X}}^{+},\mathcal{Q}^{+}),
\end{aligned}
\label{eq:pipeline}
\end{equation}
where $-$ and $+$ denote historical and future windows. $\mathcal{E}_{\theta}$ reconstructs the historical physical fields $\widehat{\mathbf{X}}^{-}$, $\mathcal{P}_{\phi}$ predicts all $T_{\mathrm{out}}$ future states $\widetilde{\mathbf{X}}^{+}$ in a single pass, $\mathcal{D}^{g}_{\psi}$ refines the full-grid forecasts, and $\mathcal{D}^{p}_{\omega}$ predicts T/S at continuous query coordinates $\mathcal{Q}^{+}$. GLORYS12 reanalysis is used during training for processor pretraining, gridded supervision of $\widehat{\mathbf{X}}^{-}$ and $\widehat{\mathbf{X}}^{+}$, and hybrid profile augmentation, but is not required at inference.

\subsection{Observation Encoder for Historical Field Reconstruction}
\label{sec:encoder}

\begin{figure*}[t]
    \centering
    \includegraphics[width=0.90\textwidth]{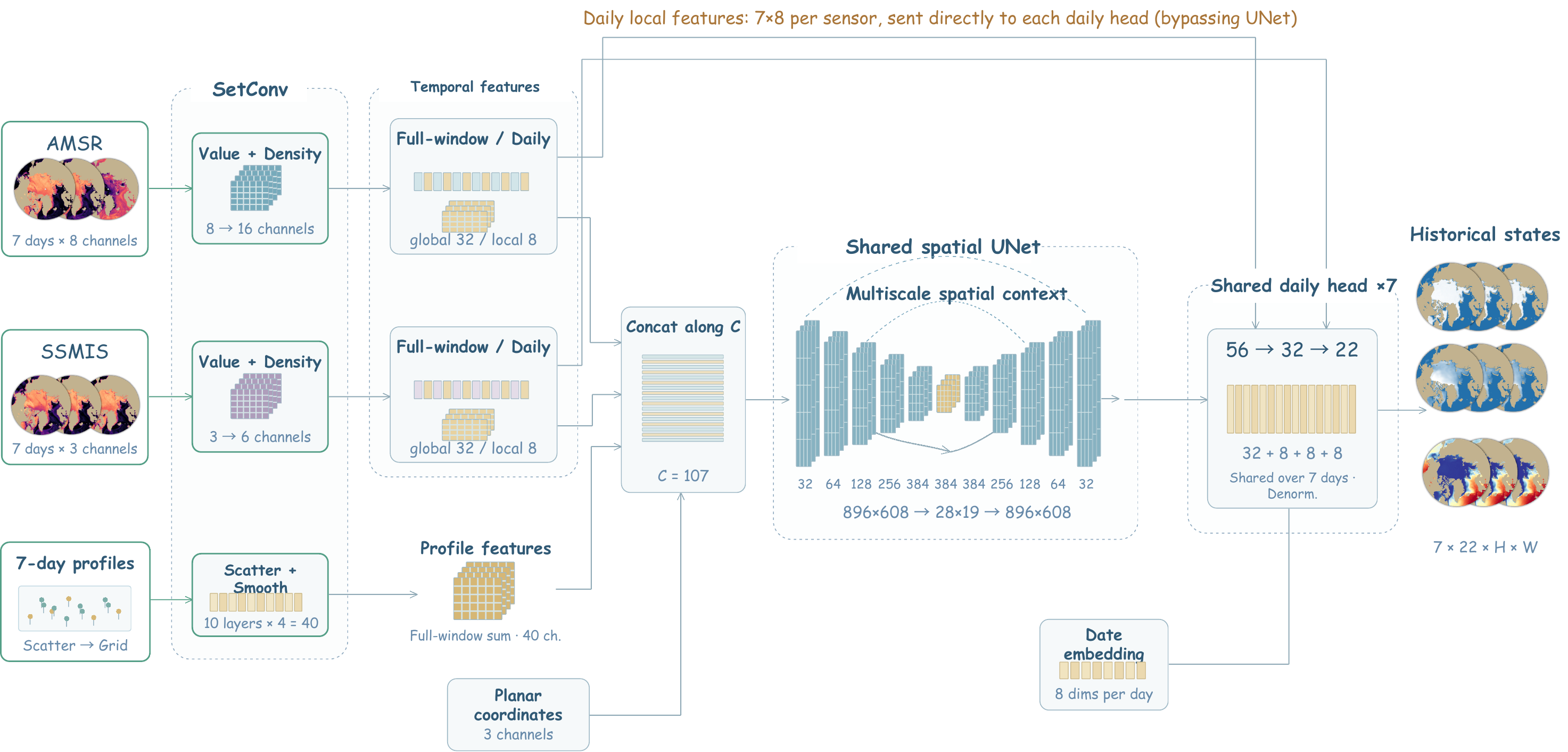}
    \caption{Observation encoder architecture. On-grid and off-grid SetConv layers extract value and density maps from AMSR, SSMIS, and profile observations. Dual-branch temporal fusion produces 32-channel window features and 8-channel daily local features per sensor. The window features, 40-channel profile features, and 3-channel planar coordinates form a 107-channel context processed once by UNetPolar, before a shared daily head ($56\to32\to22$) reconstructs the seven historical states $\widehat{\mathbf{X}}^{-}$.}
    \label{fig:encoder}
\end{figure*}

The encoder maps heterogeneous observations onto a shared polar representation and reconstructs each historical state (Fig.~\ref{fig:encoder}).

\subsubsection{Density-Aware Observation Encoding}
Satellite brightness temperatures are standardized channelwise per sensor. For valid samples $\{(\mathbf{r}_i,v_i,m_i)\}_{i=1}^{N}$ with validity $m_i\in\{0,1\}$, SetConv applies a separable 1D depthwise Gaussian kernel $k_{\sigma}$ with a learnable length scale $\sigma$ and reflective padding to compute the raw density $d_0(\mathbf{r})=\sum_{i=1}^{N} m_i k_{\sigma}(\mathbf{r}-\mathbf{r}_i)$, interpolated value $u(\mathbf{r})$, and normalized density $d(\mathbf{r})\in[0,1]$:
\begin{equation}
\begin{aligned}
    u(\mathbf{r}) &= \frac{\sum_{i=1}^{N}m_i k_{\sigma}(\mathbf{r}-\mathbf{r}_i)v_i}{d_0(\mathbf{r})+\epsilon}, \\
    d(\mathbf{r}) &= \frac{d_0(\mathbf{r})}{\max_{\mathbf{r}'}d_0(\mathbf{r}')+\epsilon}.
\end{aligned}
\label{eq:setconv}
\end{equation}
Retaining $[u,d]$ distinguishes near-zero measurements from unobserved cells. AMSR and SSMIS are processed by separate $5\times5$ on-grid SetConv branches, doubling channel counts to produce daily features $\mathbf{F}^{s}_t = \operatorname{SetConv}^{s}(\mathbf{B}^{s}_t,\mathbf{M}^{s}_t)$ ($s\in\{A,S\}$; 16 and 6 channels, respectively).

In situ ITP/WOD measurements within the upper 40~m are aligned to the nearest horizontal grid cell $(i_y,i_x)$ and the nearest of the ten nominal depth levels (retaining the observation closest to the layer center when multiple samples fall in the same layer). Each profile point is represented by eight normalized attributes: grid coordinates $(\bar y,\bar x)$, vertical layer index, continuous longitude $\bar\lambda$, latitude $\bar\varphi$, nominal layer depth $\bar z$, temperature, and salinity. In the $9\times9$ off-grid SetConv branch, valid T/S values and sample counts are accumulated onto $(i_y,i_x)$ and the corresponding depth layer, smoothed by $k_{\sigma}$, and normalized via \eqref{eq:setconv} to yield interpolated T/S fields and independent per-layer densities across the 10 levels ($\mathbf{F}^{P}\in\mathbb{R}^{40\times H\times W}$).

\subsubsection{Reliability-Aware Temporal Fusion}
For each sensor $s$, a $1\times1$ convolution (with GroupNorm and SiLU) projects $\mathbf{F}^{s}_t$ to 32-channel daily features $\mathbf{h}^{s}_t = \phi_s(\mathbf{F}^{s}_t)$. Let $\mathbf{e}^{s}_t\in\mathbb{R}^{32}$ be a learned time-step embedding and $\rho^{s}_t(\mathbf{r})\in[0,1]$ the local reliability obtained by averaging the density channels of $\mathbf{F}^{s}_t$. Using a $1\times1$ convolutional scorer $a_s$, temporal attention weights combine learned relevance with observation availability:
\begin{equation}
\begin{aligned}
    s^{s}_t(\mathbf{r}) &= a_s(\mathbf{h}^{s}_t+\mathbf{e}^{s}_t)(\mathbf{r}) + \log(\rho^{s}_t(\mathbf{r})+\epsilon), \\
    \alpha^{s}_t(\mathbf{r}) &= \frac{\exp\!\big(s^{s}_t(\mathbf{r})\big)}{\sum_{\tau}\exp\!\big(s^{s}_{\tau}(\mathbf{r})\big)}.
\end{aligned}
\label{eq:temporal_attention}
\end{equation}
We compute the attention aggregate $\overline{\mathbf{h}}^{s} = \sum_t \alpha^{s}_t\odot\mathbf{h}^{s}_t$, the reliability-weighted final frame $\mathbf{h}^{s}_{\mathrm{last}} = \rho^{s}_0\odot\mathbf{h}^{s}_0$, and the endpoint trend
\begin{equation}
    \mathbf{h}^{s}_{\mathrm{trend}} = \frac{\rho^{s}_0\odot\mathbf{h}^{s}_0 - \rho^{s}_{-T_{\mathrm{in}}+1}\odot\mathbf{h}^{s}_{-T_{\mathrm{in}}+1}}{\max(T_{\mathrm{in}}-1,1)},
\end{equation}
and project their 96-channel concatenation via a $1\times1$ conv block $f_s$ to a 32-channel window feature $\mathbf{C}^{s} = f_s([\overline{\mathbf{h}}^{s}, \mathbf{h}^{s}_{\mathrm{last}}, \mathbf{h}^{s}_{\mathrm{trend}}])$. In parallel, a $1\times1$ conv block maps $\mathbf{F}^{s}_t$ to an 8-channel daily local feature $\mathbf{L}^{s}_t$ that bypasses the spatial U-Net.

\subsubsection{Shared Spatial Context and Daily Reconstruction}
The fused satellite features, profile features, and normalized polar coordinates $\mathbf{P}_{\mathrm{coord}}=(y,x,\sqrt{x^2+y^2}/\sqrt{2})\in\mathbb{R}^{3\times H\times W}$ ($y,x\in[-1,1]$) are concatenated ($32+32+40+3=107$ channels) and processed once by UNetPolar $\mathcal{U}$ to obtain a 32-channel spatial context $\mathbf{C} = \mathcal{U}([\mathbf{C}^{A}, \mathbf{C}^{S}, \mathbf{F}^{P}, \mathbf{P}_{\mathrm{coord}}])$. UNetPolar employs five stride-2 downsampling stages ($32\to64\to128\to256\to384$, reaching a $28\times19$ bottleneck) and symmetric bilinear upsampling stages with skip connections and reflective padding. A shared two-layer $1\times1$ conv head $f_{\mathrm{hist}}$ ($56\to32\to22$, GroupNorm and SiLU) combines $\mathbf{C}$ with the daily local features and an 8-dimensional time embedding $\mathbf{e}^{\mathrm{hist}}_t$ to reconstruct each historical state:
\begin{equation}
    \widehat{\mathbf{X}}_t = \mathcal{N}^{-1}_{X}\!\big( f_{\mathrm{hist}}([\mathbf{C}, \mathbf{L}^{A}_t, \mathbf{L}^{S}_t, \mathbf{e}^{\mathrm{hist}}_t]) \big),
    \label{eq:daily_reconstruction}
\end{equation}
where $\mathcal{N}^{-1}_{X}$ applies channelwise inverse standardization to restore physical units.

\subsection{Temporal Processor for Physical-State Forecasting}
\label{sec:processor}

The Swin-based processor predicts all $T_{\mathrm{out}}=7$ future states jointly from $\widehat{\mathbf{X}}^{-}$ without autoregressive rollout:
\begin{equation}
    \widetilde{\mathbf{X}}^{+} = \mathcal{N}^{-1}_{X}\!\big[ \mathcal{R}_{\phi}\!\big( \mathcal{T}_{\phi}\!\big( \mathcal{S}_{\phi}(\mathcal{N}_{X}(\widehat{\mathbf{X}}^{-})) \big) \big) \big],
    \label{eq:swin_processor}
\end{equation}
where $\mathcal{N}_{X}$ is a learnable channelwise standardization layer initialized from training statistics. The spatial encoder $\mathcal{S}_{\phi}$ applies four $3\times3$ conv blocks ($C_{\mathrm{hid}}=16$) independently to each historical frame, downsampling by stride 2 at the second and fourth blocks ($896\times608\to224\times152$) while retaining the first-block features as a skip connection. The translator $\mathcal{T}_{\phi}$ flattens the time and channel axes into $T_{\mathrm{in}}C_{\mathrm{hid}}=112$ channels and processes them through four Swin Transformer blocks ($1\times1$ projections $112\to256\to112$, 8 heads, window size $7\times7$, MLP ratio 4) alternating regular and cyclically shifted window partitions, with window attention
\begin{equation}
    \operatorname{Attn}(\mathbf{Q},\mathbf{K},\mathbf{V}) = \operatorname{softmax}\!\left( \frac{\mathbf{Q}\mathbf{K}^{\mathsf{T}}}{\sqrt{d}} +\mathbf{B}_{\mathrm{rel}} \right)\mathbf{V},
    \label{eq:window_attention}
\end{equation}
where $d$ is the head dimension and $\mathbf{B}_{\mathrm{rel}}$ is the relative positional bias. Reshaping the 112-channel output back to $(T_{\mathrm{out}},C_{\mathrm{hid}},224,152)$ and passing it through the symmetric four-layer spatial decoder $\mathcal{R}_{\phi}$ (with PixelShuffle $\times2$ upsampling and the encoder skip connection) yields $\widetilde{\mathbf{X}}^{+}$.

\subsection{Decoder for Grid Refinement and Point Queries}
\label{sec:decoder}

\begin{figure*}[t]
    \centering
    \includegraphics[width=0.90\textwidth]{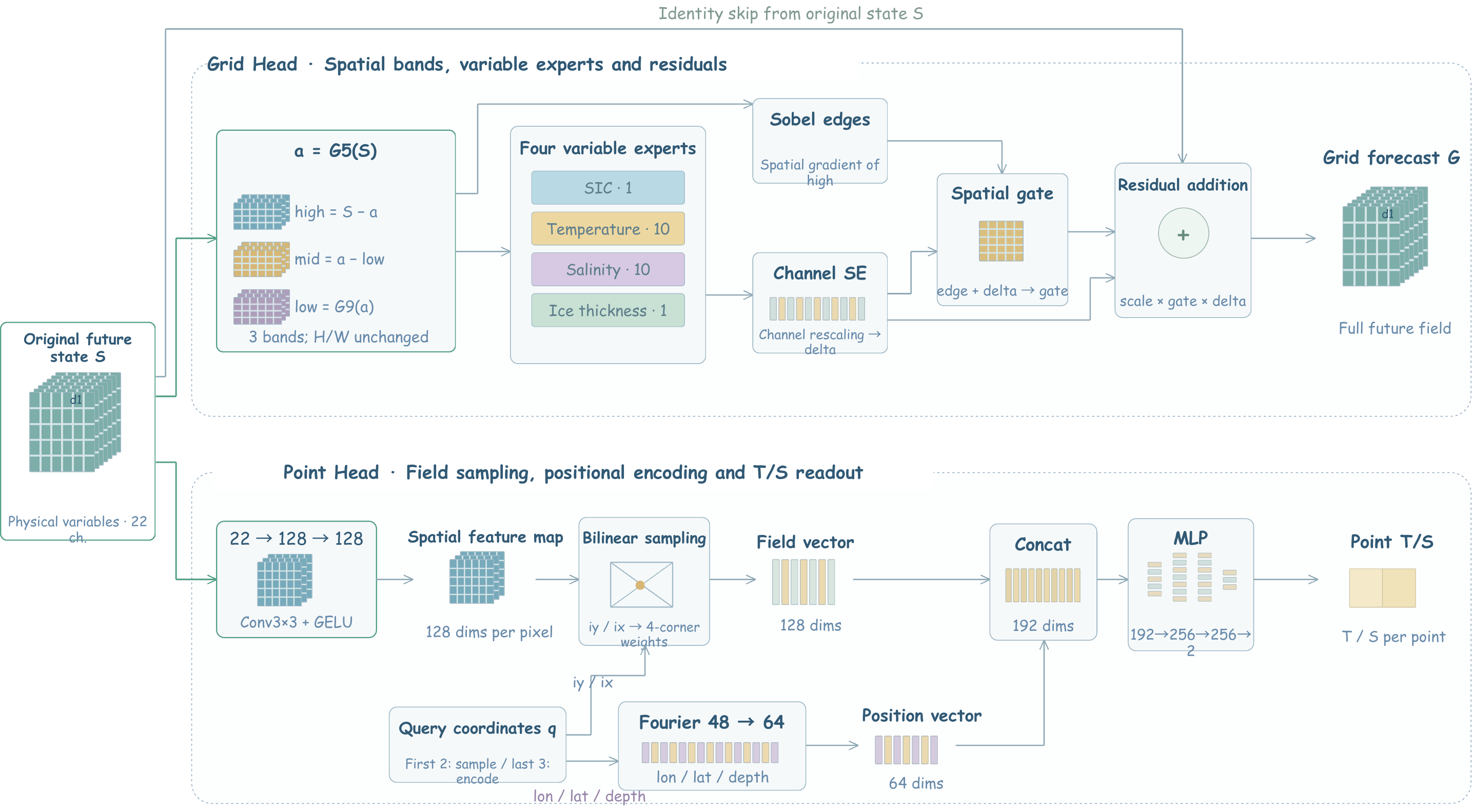}
    \caption{Decoder architecture operating on $\widetilde{\mathbf{X}}_k$. \emph{Top:} the gridded head decomposes the state via cascaded $5\times5$ and $9\times9$ separable smoothing filters into fine-, intermediate-, and coarse-scale spatial components, processes them through four variable-group experts (width 48), and injects the channel-recalibrated residual via a Sobel edge-guided spatial gate. \emph{Bottom:} the point head bilinearly samples a 128-channel feature map at $(\bar y,\bar x)$, concatenates a 64-dimensional sinusoidal position embedding of $(\bar\lambda,\bar\varphi,\bar z)$, and predicts T/S via an MLP.}
    \label{fig:decoder}
\end{figure*}

At each forecast lead $k$, parallel gridded and pointwise branches decode the processor output $\mathbf{Z}=\widetilde{\mathbf{X}}_k\in\mathbb{R}^{C\times H\times W}$ with shared weights across leads (Fig.~\ref{fig:decoder}).

\subsubsection{Multiscale Spatial-Difference Grid Refinement}
Two cascaded channelwise separable spatial smoothing filters, $\mathcal{B}_5$ ($5\times1$ and $1\times5$ depthwise convolutions with reflective padding) and $\mathcal{B}_9$ ($9\times1$ and $1\times9$), initialized with unit-variance 1D Gaussian weights and kept learnable during training, decompose $\mathbf{Z}$ by successive spatial subtraction into three full-resolution spatial components summing identically to $\mathbf{Z}$:
\begin{equation}
\begin{aligned}
    \mathbf{Z}_{\mathrm{coarse}} &= \mathcal{B}_9(\mathcal{B}_5(\mathbf{Z})), \\
    \mathbf{Z}_{\mathrm{mid}} &= \mathcal{B}_5(\mathbf{Z})-\mathbf{Z}_{\mathrm{coarse}}, \quad
    \mathbf{Z}_{\mathrm{fine}} = \mathbf{Z}-\mathcal{B}_5(\mathbf{Z}).
\end{aligned}
\label{eq:band_decomposition}
\end{equation}
The 22 channels are split into four physical variable groups $g\in\{1,2,3,4\}$ (SIC, 10-level temperature, 10-level salinity, SIT; $C_g\in\{1,10,10,1\}$). Within expert $g$, each spatial component $\mathbf{Z}^{g}_{s}$ ($s\in\{\mathrm{fine},\mathrm{mid},\mathrm{coarse}\}$) is projected to $W_e=48$ channels by a $1\times1$ conv and processed by scale-specific spatial blocks $\mathbf{H}^{g}_{s}=\psi^{(g)}_{s}(\mathbf{Z}^{g}_{s})$: $\psi^{(g)}_{\mathrm{fine}}$ applies parallel $1\times7$ and $7\times1$ anisotropic depthwise convs (concatenated and mixed to 48 channels via $1\times1$ conv and GELU) followed by a $3\times3$ depthwise conv, GELU, and $1\times1$ projection; $\psi^{(g)}_{\mathrm{mid}}$ applies a $5\times5$ depthwise conv, GELU, and $1\times1$ projection; and $\psi^{(g)}_{\mathrm{coarse}}$ applies a $1\times1$ conv and GELU. The three 48-channel outputs are concatenated (144 channels) and projected back to $C_g$ channels via a $1\times1$ conv $W^{(g)}_{\mathrm{out}}$:
\begin{equation}
    \boldsymbol{\Delta}_g = W^{(g)}_{\mathrm{out}} * \big[ \mathbf{H}^{g}_{\mathrm{fine}},\, \mathbf{H}^{g}_{\mathrm{mid}},\, \mathbf{H}^{g}_{\mathrm{coarse}} \big] \in\mathbb{R}^{C_g\times H\times W}.
    \label{eq:scale_recombination}
\end{equation}

Concatenating all group residuals into $\boldsymbol{\Delta}=\operatorname{Concat}_{g=1}^{4}(\boldsymbol{\Delta}_g)\in\mathbb{R}^{22\times H\times W}$, we apply squeeze-and-excitation channel recalibration $\boldsymbol{\Delta}' = \boldsymbol{\Delta}\odot\sigma(f_{\mathrm{ch}}(\operatorname{GAP}(\boldsymbol{\Delta})))$ with a $22\to16\to22$ bottleneck $f_{\mathrm{ch}}$. To sharpen ice edges and oceanic fronts, a fixed $3\times3$ Sobel operator extracts the spatial gradient magnitude $\mathbf{E}\in\mathbb{R}^{1\times H\times W}$ of the channel-averaged fine-scale component $\mathbf{Z}_{\mathrm{fine}}$, which is projected to 22 channels by a $1\times1$ conv $f_{\mathrm{edge}}$ and combined with $\boldsymbol{\Delta}'$ inside a $3\times3$ depthwise conv spatial gate $f_{\mathrm{dw}}$:
\begin{equation}
\begin{aligned}
    \mathbf{G}_{\mathrm{spatial}} &= \sigma\!\big(f_{\mathrm{dw}}\big(f_{\mathrm{edge}}(\mathbf{E}) + \boldsymbol{\Delta}'\big)\big), \\
    \widehat{\mathbf{X}}_k &= \mathbf{Z} + \eta\,\mathbf{G}_{\mathrm{spatial}}\odot\boldsymbol{\Delta}',
\end{aligned}
\label{eq:refined_grid}
\end{equation}
where $\eta$ is a learnable scalar initialized to $0.1$.

\subsubsection{Coordinate-Conditioned T/S Queries}
For a normalized query $\mathbf{q}=(\bar y,\bar x,\bar\lambda,\bar\varphi,\bar z)$ with grid coordinates $\bar y,\bar x\in[0,1]$ and geographic coordinates $\mathbf{g}=(\bar\lambda,\bar\varphi,\bar z)\in\mathbb{R}^{3}$, two $3\times3$ conv layers ($22\to128\to128$, GELU) extract a feature map $\mathbf{F}_k=f_{\mathrm{point}}(\widetilde{\mathbf{X}}_k)$, which is bilinearly sampled at $(\bar y,\bar x)$ to obtain $\mathbf{f}_{k,\mathbf{q}}\in\mathbb{R}^{128}$. Simultaneously, $\mathbf{g}$ is encoded via $J=8$ octave sinusoidal frequencies and linearly projected by $f_{\mathrm{geo}}$ to a 64-dimensional position embedding, before a three-layer MLP ($192\to256\to256\to2$, GELU) predicts normalized T/S:
\begin{equation}
\begin{aligned}
    \gamma(\mathbf{g}) &= \operatorname{Concat}_{j=0}^{J-1}\!\big[\sin(2^j\mathbf{g}),\,\cos(2^j\mathbf{g})\big]\in\mathbb{R}^{48}, \\
    \widehat{\mathbf{y}}_{k,\mathbf{q}} &= f_{\mathrm{MLP}}\!\big([\mathbf{f}_{k,\mathbf{q}},\, f_{\mathrm{geo}}(\gamma(\mathbf{g}))]\big)\in\mathbb{R}^{2},
\end{aligned}
\label{eq:point_prediction}
\end{equation}
followed by inverse standardization to physical units ($^{\circ}\mathrm{C}$ and $\mathrm{psu}$). Operating on $\widetilde{\mathbf{X}}_k$ in parallel with $\mathcal{D}^{g}_{\psi}$ allows $\mathcal{D}^{p}_{\omega}$ to adapt directly to in situ supervision.

\subsection{Training Objectives and Optimization}
\label{sec:training}

To balance variables with distinct units and variances across the four variable groups $\mathcal{G}$ (channel index sets $\mathcal{C}_g$), per-channel weights $w_c = \frac{1}{|\mathcal{G}|} \big(\sigma_c^{-2} / \sum_{j\in\mathcal{C}_g}\sigma_j^{-2}\big)$ ($c\in\mathcal{C}_g$) normalize the inverse training variance $\sigma_c^{-2}$ within each group so that each group accounts for $25\%$ of the total weight. For predicted and reference sequences $\mathbf{U}$ and $\mathbf{V}$ under channelwise ocean masks $\mathbf{M}$, the group-balanced masked RMSE is
\begin{equation}
    \mathcal{R}(\mathbf{U},\mathbf{V};\mathbf{M}) = \sqrt{ \sum_{c=1}^{C} w_c\, \frac{ \sum_{t,\mathbf{r}} M_{t,c,\mathbf{r}} (U_{t,c,\mathbf{r}}-V_{t,c,\mathbf{r}})^2 }{ \sum_{t,\mathbf{r}}M_{t,c,\mathbf{r}}+\epsilon } +\epsilon }.
    \label{eq:masked_rmse}
\end{equation}
Historical reconstruction and future gridded losses against GLORYS12 ($\mathrm{G}$) are $\mathcal{L}_{\mathrm{hist}} = \mathcal{R}(\widehat{\mathbf{X}}^{-}, \mathbf{X}^{-,\mathrm{G}}; \mathbf{M}^{-})$ and $\mathcal{L}_{\mathrm{grid}} = \mathcal{R}(\widehat{\mathbf{X}}^{+}, \mathbf{X}^{+,\mathrm{G}}; \mathbf{M}^{+})$. To preserve sharp ice edges and fronts, a spatial-gradient loss penalizes first-order horizontal and vertical differences over valid neighboring cell pairs $\langle\cdot\rangle_{\mathbf{M}_c}$:
\begin{equation}
\begin{split}
    \mathcal{L}_{\mathrm{grad}} = \sum_{c=1}^{C} w_c\, \Big\langle
    &\big| \nabla_x\widehat{\mathbf{X}}^{+}_c - \nabla_x\mathbf{X}^{+,\mathrm{G}}_c \big|, \\
    &\big| \nabla_y\widehat{\mathbf{X}}^{+}_c - \nabla_y\mathbf{X}^{+,\mathrm{G}}_c \big|
    \Big\rangle_{\mathbf{M}_c}.
\end{split}
\label{eq:gradient_loss}
\end{equation}
For future in situ observations in normalized T/S space with validity mask $m^p$, the pointwise MSE and total objective are
\begin{equation}
\begin{aligned}
    \mathcal{L}_{\mathrm{point}} &= \frac{ \sum_{k,q,c} m^{p}_{k,q,c} (\widehat{y}_{k,q,c}-y^{\mathrm{obs}}_{k,q,c})^2 }{ \sum_{k,q,c}m^{p}_{k,q,c}+\epsilon }, \\
    \mathcal{L} &= \mathcal{L}_{\mathrm{grid}} + 0.2\mathcal{L}_{\mathrm{grad}} + 0.5\mathcal{L}_{\mathrm{hist}} + \mathcal{L}_{\mathrm{point}}.
\end{aligned}
\label{eq:total_loss}
\end{equation}

We first pretrain $\mathcal{P}_{\phi}$ on GLORYS12 reanalysis-to-reanalysis forecasting, $\mathcal{E}_{\theta}$ on historical field reconstruction, and $\mathcal{D}_{\psi,\omega}$ with $\mathcal{E}_{\theta}$ and $\mathcal{P}_{\phi}$ frozen, and then fine-tune all three modules end-to-end in bfloat16 mixed precision (preventing gradient overflow in sparse SetConv density normalization) using AdamW (base learning rate $10^{-4}$, cosine schedule, 3 warm-up epochs, gradient clipping at $1.0$, module learning-rate multipliers $1.0/0.1/0.5$ for $\mathcal{E}_{\theta}/\mathcal{P}_{\phi}/\mathcal{D}_{\psi,\omega}$). During training, we apply two window-level profile regularizers: with $50\%$ probability, both historical input profiles and future pointwise target values within a window are replaced by GLORYS12 T/S values sampled at the real profile coordinates and layers plus Gaussian noise ($\sigma=0.05$); independently, historical profile inputs undergo a $70\%$ window-level modality dropout to simulate operational profile absence. Validation uses $100\%$ real in situ profile inputs and targets without dropout.

\section{Experiments}
\label{sec:experiments}

We evaluate \Ours in terms of multivariate forecast accuracy, spatial and vertical forecast structure, sensitivity to model components and observation availability, and performance across seasonal regimes and extreme sea-ice events.

\subsection{Experimental Setup}
\label{sec:exp_setup}

The experiments combine AMSR and SSMIS brightness temperatures with ITP/WOD T/S profiles to predict GLORYS12 physical fields on a common Arctic grid. The grid contains $896\times608$ cells at approximately 12-km resolution. Each sample uses seven historical days to predict the following seven days. Outputs comprise SIC, SIT, and potential temperature and salinity at ten nominal depths between 0 and 40~m, yielding 22 physical channels. Nominal depths are mapped to the nearest available reanalysis levels, and depth-resolved evaluation uses the corresponding source coordinates.

The training period spans 2006--2022 and provides 5,723 valid windows. The validation set contains 961 daily initializations from January 7, 2023, to August 24, 2025, with forecast verification through August 31, 2025, and is used for checkpoint selection and evaluation. Gridded metrics exclude land and invalid reference values; SIT evaluation includes zero-thickness open-water cells within the valid ocean mask. The implementation uses PyTorch, and the reported evaluation is performed on a single NVIDIA H200 NVL GPU.

\subsection{Evaluation Metrics}
\label{sec:evaluation_metrics}

We evaluate gridded forecasts against GLORYS12 reanalysis and pointwise predictions against matched in situ T/S observations. The RMSE, mean absolute error (MAE), and signed bias are defined as
\begin{align}
    \mathrm{RMSE} &= \sqrt{\frac{\sum_{i\in\Omega} w_i(\widehat{y}_i-y_i)^2}{\sum_{i\in\Omega} w_i}}, \label{eq:eval_rmse}\\
    \mathrm{MAE} &= \frac{\sum_{i\in\Omega} w_i|\widehat{y}_i-y_i|}{\sum_{i\in\Omega} w_i}, \label{eq:eval_mae}\\
    \mathrm{Bias} &= \frac{\sum_{i\in\Omega} w_i(\widehat{y}_i-y_i)}{\sum_{i\in\Omega} w_i}, \label{eq:eval_bias}
\end{align}
where $\Omega$ contains valid prediction--reference pairs, $\widehat{y}_i$ and $y_i$ denote the prediction and reference, and $w_i$ is the evaluation weight. SIC errors are reported in percentage points (pp), potential-temperature errors in $^\circ$C, salinity errors in $10^{-3}$, and SIT errors in m.

For the observation-to-forecast model comparison, valid pixels are equally weighted. Channel-wise metrics are computed by pooling D1--D7 within each initialization window and then averaging over the validation windows. For the spatial, lead-time, refinement, and seasonal analyses, metrics are computed at each lead by pooling valid grid cells across the relevant initialization windows with grid-cell area weights. Cell areas are approximated by the product of local horizontal and vertical distances between adjacent grid centers on the WGS84 ellipsoid. Temperature and salinity scores are averaged equally over ten depth levels; seven-day summaries of the area-weighted results average the seven lead-specific scores. The archived training ablations use pixel-weighted metrics.

Pointwise evaluation assigns equal weight to each matched forecast--observation sample. The point head and trilinear interpolation of the refined grid are evaluated on identical samples within the common depth interval of $[0.494025,40]$~m, without vertical extrapolation. Their seven-day scores are averages of the lead-specific metrics.

Sea-ice extent and integrated ice-edge error (IIEE) are calculated using a SIC threshold of 15\%. For grid-cell area $a_j$ and SIC expressed as a fraction,
\begin{align}
    \mathrm{SIE}(c) &= \sum_j a_j\,\mathbb{I}(c_j\geq 0.15), \label{eq:eval_sie}\\
    \mathrm{IIEE}(\widehat{c},c) &= \sum_j a_j\left|\mathbb{I}(\widehat{c}_j\geq 0.15)-\mathbb{I}(c_j\geq 0.15)\right|, \label{eq:eval_iiee}
\end{align}
where $\mathbb{I}(\cdot)$ is the indicator function. The signed extent error is $\mathrm{SIE}(\widehat{c})-\mathrm{SIE}(c)$. The marginal ice zone (MIZ) is defined by reference SIC between 15\% and 80\%. Spatial-gradient errors are evaluated using first-order differences between adjacent valid grid cells, normalized by their geodesic separation, and summarized as area-weighted absolute errors.

For an error metric $E$ and a positive baseline error $E_{\mathrm{base}}$, relative error reduction is $100(E_{\mathrm{base}}-E_{\mathrm{model}})/E_{\mathrm{base}}$. Positive values indicate lower forecast error.

\subsection{Training Configuration}
\label{sec:exp_training}

Arctic E2E follows the staged pretraining and joint optimization procedure introduced in Section~\ref{sec:training}. Joint fine-tuning uses a batch size of one, AdamW weight decay of 0.05, and bfloat16 mixed precision. The final fine-tuning run lasts five epochs, and the checkpoint with the lowest validation objective is selected. Historical ITP/WOD observations provide profile inputs, while future profiles supply pointwise targets. Each historical window contains at most 512 profile tokens.

\subsection{Comparison with Baseline Methods}
\label{sec:exp_comparison}

We compare Arctic E2E with six observation-to-forecast baselines: ConvLSTM, SimVP, TAU, ConvNeXt, Swin Transformer, and InceptionNeXt. All methods use AMSR and SSMIS observations together with historical ITP/WOD profiles, without historical reanalysis fields as forecasting inputs. Arctic E2E represents profiles as point tokens, whereas the baselines use daily sparse grids obtained by averaging observations within each grid cell and depth layer. All methods are evaluated on the same validation windows using the pixel-weighted, window-pooled protocol defined in Section~\ref{sec:evaluation_metrics}.

Table~\ref{tab:exp_comparison} shows that Arctic E2E achieves the best performance in six of the eight evaluation metrics, ranking first in both RMSE and MAE for SIC, SIT, and salinity. Its largest improvements occur in sea-ice prediction. For SIC, Arctic E2E obtains an RMSE of 8.045~pp and an MAE of 2.540~pp, reducing the errors of ConvLSTM, the strongest baseline for both metrics, by 32.0\% and 56.2\%, respectively. The improvement in both metrics indicates that the gain extends beyond a single error criterion.

Arctic E2E also provides the most accurate SIT forecasts. Its RMSE is 0.162~m, 22.3\% lower than that of SimVP, the strongest baseline for this metric. Its MAE is 0.062~m, improving upon the best baseline MAE, obtained by ConvNeXt, by 42.7\%. Together, the SIC and SIT results demonstrate the effectiveness of the proposed framework for jointly predicting sea-ice coverage and thickness directly from observations.

For salinity, Arctic E2E achieves an RMSE of $0.508\times10^{-3}$ and an MAE of $0.297\times10^{-3}$, improving upon ConvNeXt by 2.3\% and 3.3\%, respectively. ConvNeXt performs best for potential temperature, with RMSE and MAE of 0.840~$^\circ$C and 0.553~$^\circ$C, compared with 0.919~$^\circ$C and 0.585~$^\circ$C for Arctic E2E. Overall, Arctic E2E combines substantial sea-ice improvements with the lowest salinity errors in a single observation-driven forecasting system.

\begin{table*}[t]
    \centering
    \caption{Observation-to-forecast performance over the seven-day forecast horizon. Scores use equally weighted valid pixels and window-pooled aggregation. Units are pp for SIC, $^\circ$C for potential temperature, $10^{-3}$ for salinity, and m for SIT. Lower values are better. The best and second-best results in each column are bold and underlined, respectively.}
    \label{tab:exp_comparison}
    \small
    \setlength{\tabcolsep}{5pt}
    \renewcommand{\arraystretch}{1.12}
    \begin{tabular}{lrrrrrrrr}
        \toprule
        & \multicolumn{4}{c}{RMSE $\downarrow$} & \multicolumn{4}{c}{MAE $\downarrow$} \\
        \cmidrule(lr){2-5}\cmidrule(lr){6-9}
        Method & SIC & Temp. & Salinity & SIT & SIC & Temp. & Salinity & SIT \\
        \midrule
        ConvLSTM & \underline{11.827} & 1.209 & 1.142 & 0.236 & \underline{5.802} & 0.806 & 0.654 & 0.124 \\
        SimVP & 12.924 & 0.881 & 0.528 & \underline{0.208} & 7.076 & \underline{0.575} & 0.312 & 0.110 \\
        TAU & 15.347 & 1.682 & 0.814 & 0.260 & 8.849 & 1.302 & 0.498 & 0.132 \\
        ConvNeXt & 13.378 & \textbf{0.840} & \underline{0.520} & 0.210 & 7.861 & \textbf{0.553} & \underline{0.307} & \underline{0.108} \\
        Swin Transformer & 13.301 & \underline{0.874} & 0.549 & 0.229 & 7.624 & 0.580 & 0.322 & 0.123 \\
        InceptionNeXt & 14.345 & 1.698 & 0.924 & 0.267 & 8.144 & 1.258 & 0.543 & 0.153 \\
        \midrule
        Arctic E2E & \textbf{8.045} & 0.919 & \textbf{0.508} & \textbf{0.162} & \textbf{2.540} & 0.585 & \textbf{0.297} & \textbf{0.062} \\
        \bottomrule
    \end{tabular}
\end{table*}

\subsection{Spatial, Vertical, and Pointwise Forecast Performance}
\label{sec:exp_forecast}

\subsubsection{Spatial Forecasts and Lead-Time Dependence}

\begin{figure}[t]
    \centering
    \includegraphics[width=0.9\columnwidth]{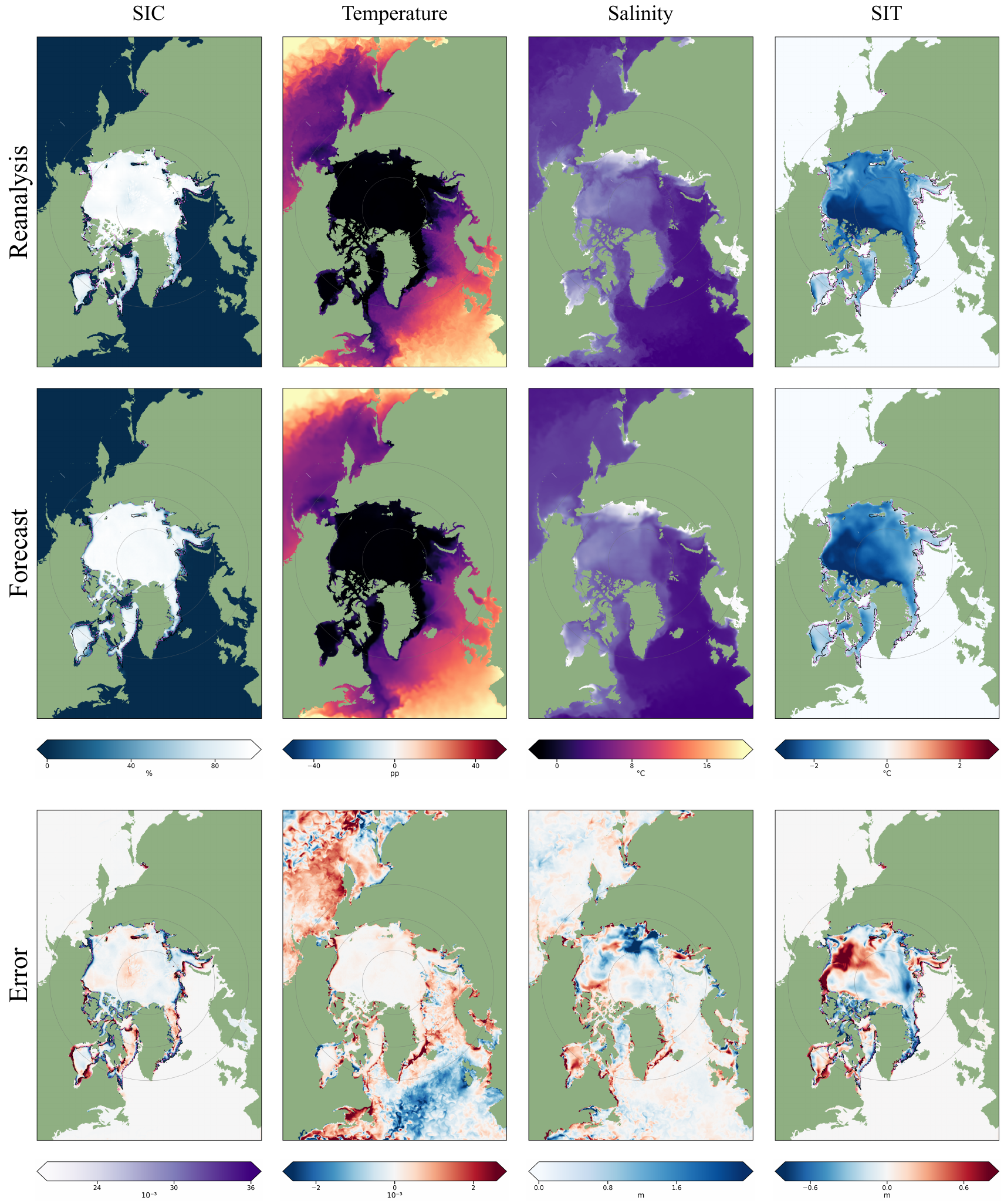}
    \caption{D7 spatial forecasts initialized on May 31, 2023, and verified on June 7, 2023. Columns show SIC, potential temperature, salinity, and SIT; rows show GLORYS12 reanalysis, Arctic E2E forecasts, and forecast-minus-reanalysis errors. Temperature and salinity are shown at the 0.494-m source level. Black solid and magenta dashed contours indicate the reference and predicted 15\% SIC ice edges. Reference and forecast fields share color scales. Display limits use the reference 2nd--98th percentiles for fields and the 98th percentile of absolute errors for symmetric error scales.}
    \label{fig:spatial_forecast}
\end{figure}

Beyond aggregate errors, Arctic E2E produces complete sea-ice and upper-ocean fields over the seven-day forecast horizon. Fig.~\ref{fig:spatial_forecast} presents a D7 forecast initialized on May 31, 2023, and verified on June 7, 2023. The model reproduces the broad distribution of ice-covered and open-water regions, the large-scale SIT pattern, and the main spatial contrasts in near-surface temperature and salinity. These results illustrate the ability of the unified model to generate multiple physically interpretable fields from the same historical observation window.

The error maps further distinguish the spatial characteristics of the predicted variables. SIC differences are concentrated near the ice edge, where displacement of the predicted boundary produces localized errors. SIT differences extend into the central ice pack, whereas temperature and salinity errors exhibit coherent regional patterns. The forecasts preserve the large-scale structures while smoothing some of the sharper transitions present in the reanalysis.

Area-weighted, lead-specific metrics show gradual error growth over the forecast horizon. From D1 to D7, SIC RMSE increases from 8.241 to 9.038~pp, temperature RMSE from 0.874 to 0.902~$^\circ$C, salinity RMSE from $0.532$ to $0.536\times10^{-3}$, and SIT RMSE from 0.173 to 0.178~m. To quantify the contribution of forecasting beyond observation-based reconstruction, we compare against encoder persistence, which holds the final reconstructed historical field fixed. At D7, Arctic E2E reduces RMSE by 5.0\%, 7.4\%, 1.8\%, and 4.6\% for SIC, temperature, salinity, and SIT, respectively. For SIC, the advantage emerges at D5--D7, following higher RMSE than persistence at D1--D4. The D7 improvements across all four variables demonstrate that the learned forecast evolution adds predictive value to the reconstructed initial state.

\subsubsection{Vertical Structure and Coordinate-Conditioned Predictions}

\begin{figure}[t]
    \centering
    \includegraphics[width=0.9\columnwidth]{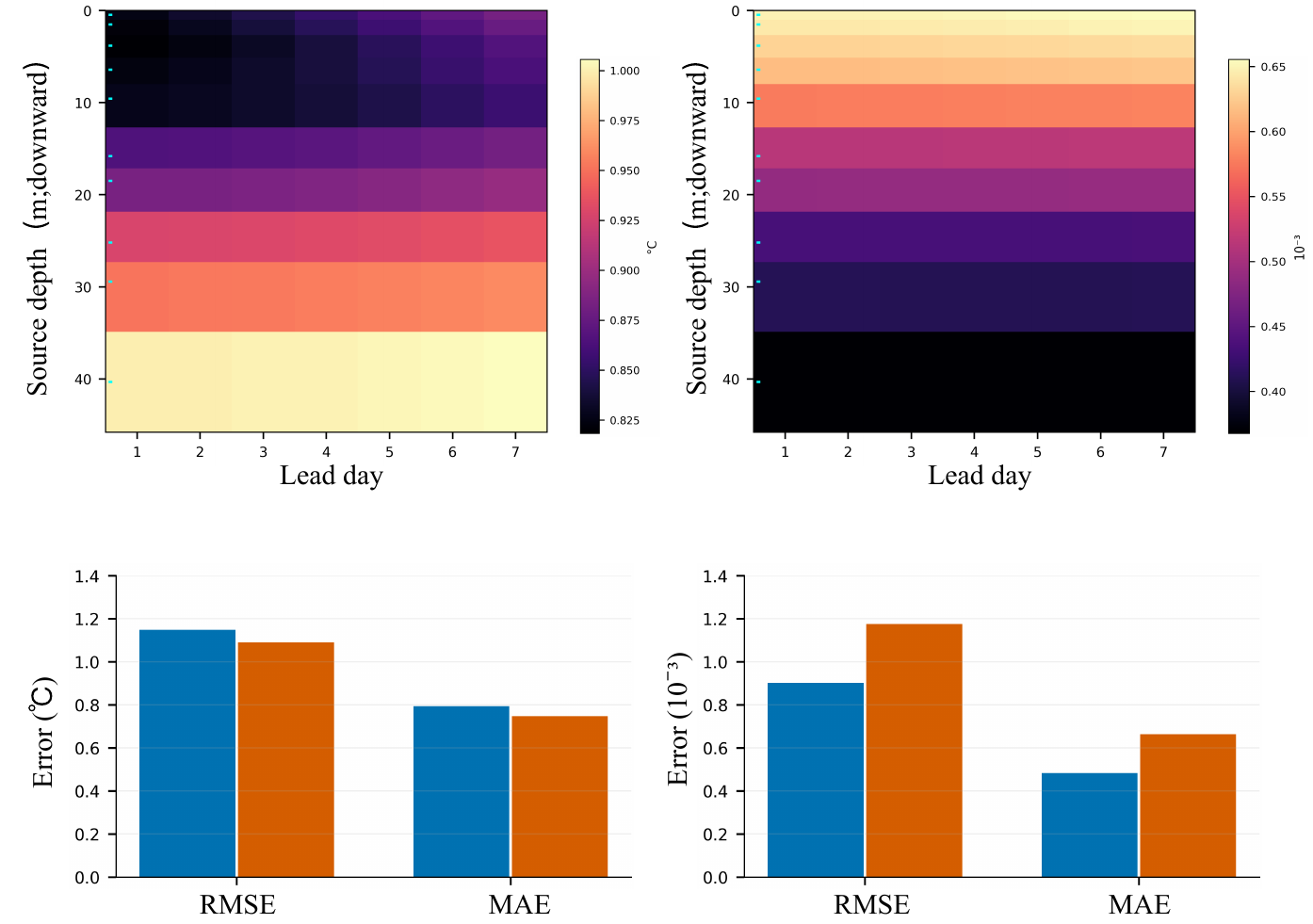}
    \caption{Depth-resolved and pointwise forecast performance. The upper panels show area-weighted potential-temperature and salinity RMSE as functions of forecast lead and source depth. The lower panels compare the point head (blue) with trilinear interpolation of the refined grid (orange), using identical matched samples. Pointwise scores average the seven lead-specific metrics. Depth increases downward, and cyan marks indicate the actual source levels.}
    \label{fig:depth_point}
\end{figure}

Arctic E2E extends the gridded forecast from surface sea ice to the vertical structure of the upper ocean. Fig.~\ref{fig:depth_point} summarizes temperature and salinity RMSE across source depths and forecast leads. The two variables exhibit distinct vertical error distributions that remain broadly consistent across D1--D7. At D7, temperature RMSE decreases from 0.883~$^\circ$C at 0.494~m to 0.856~$^\circ$C near 9.573~m before increasing to 1.006~$^\circ$C at 40.344~m. Salinity RMSE decreases with depth, from $0.655\times10^{-3}$ to $0.369\times10^{-3}$ across the same vertical range.

The coordinate-conditioned branch additionally provides predictions at observation locations and depths. Evaluated on 346,982 matched forecast--observation samples accumulated across initialization windows and leads, the point head reduces seven-day mean salinity RMSE from $1.175\times10^{-3}$ for refined-grid interpolation to $0.902\times10^{-3}$, an improvement of 23.2\%. Salinity MAE decreases from $0.663\times10^{-3}$ to $0.483\times10^{-3}$. For temperature, grid interpolation yields lower RMSE and MAE, at 1.091~$^\circ$C and 0.748~$^\circ$C, compared with 1.148~$^\circ$C and 0.794~$^\circ$C for the point head. The salinity results demonstrate the benefit of learning a location- and depth-conditioned readout in addition to the regular-grid output.

\subsection{Ablation Studies}
\label{sec:exp_ablation}

We examine how grid refinement, staged training, temporal-window design, and observation encoding contribute to forecast performance. Refinement and inference-time profile removal use fixed model weights and area-weighted evaluation. Training and architecture variants use their respective retraining configurations and pixel-weighted evaluation.

\subsubsection{Contribution of Grid Refinement}

Grid refinement provides the largest directly measured component-level improvement. As shown in Table~\ref{tab:exp_refinement}, refinement reduces seven-day mean SIC RMSE from 14.541 to 8.610~pp, temperature RMSE from 1.126 to 0.887~$^\circ$C, and salinity RMSE from $0.575$ to $0.534\times10^{-3}$. These changes correspond to reductions of 40.8\%, 21.2\%, and 7.2\%, respectively. MAE decreases for all four variables, by 68.4\% for SIC, 36.3\% for temperature, 19.8\% for salinity, and 26.7\% for SIT.

\begin{figure}[t]
    \centering
    \includegraphics[width=0.9\columnwidth]{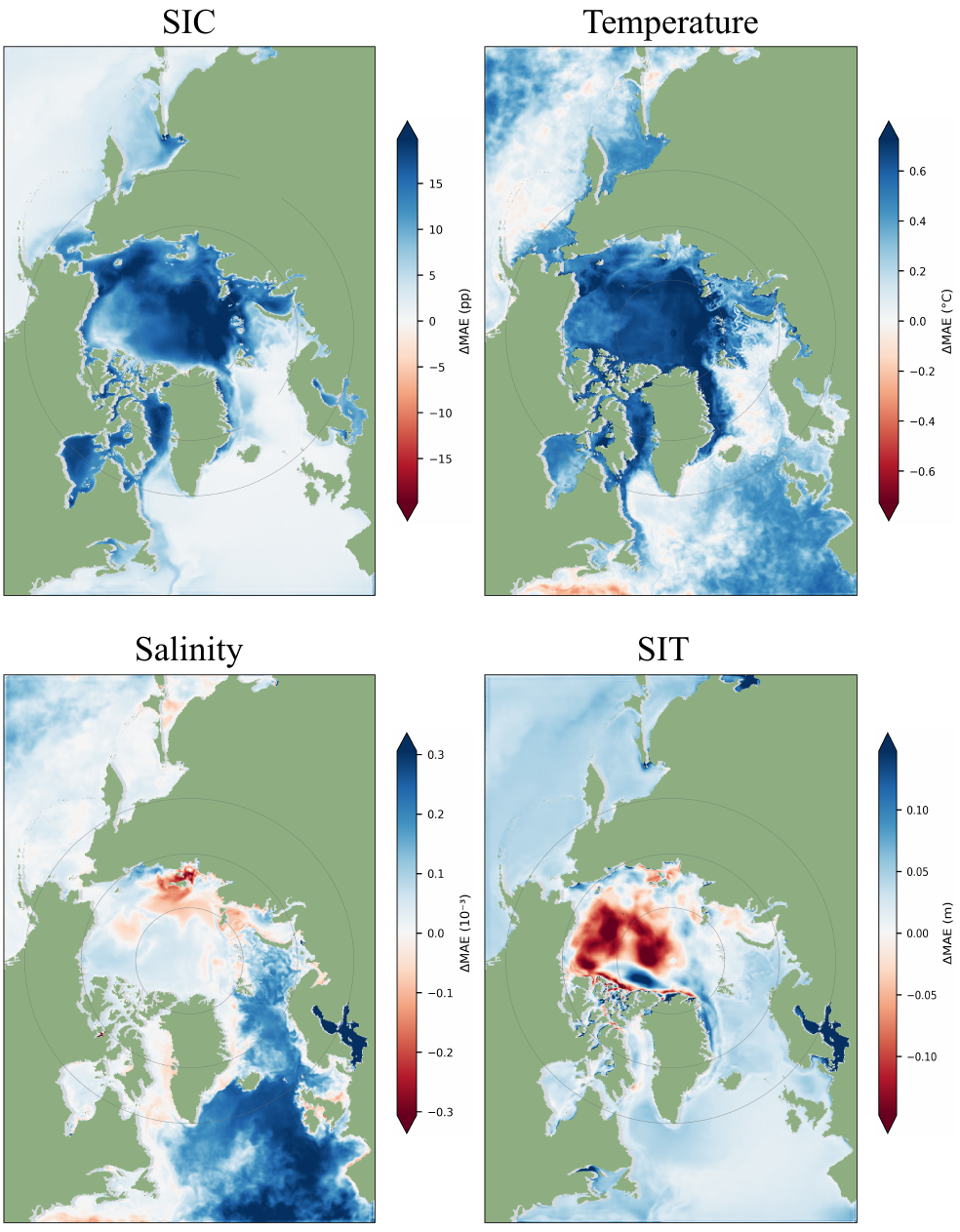}
    \caption{Spatial changes in MAE produced by grid refinement. Panels show SIC, potential temperature, salinity, and SIT in reading order. The plotted quantity is pre-refinement MAE minus refined MAE, accumulated over all 961 initialization windows, D1--D7, and the available depths of each variable group. Blue indicates improvement and red indicates deterioration. Color limits are symmetric at the 98th percentile of the absolute changes.}
    \label{fig:refinement_gain}
\end{figure}

Fig.~\ref{fig:refinement_gain} shows that these gains extend across broad spatial regions, particularly for SIC and temperature. Salinity improvements are more spatially heterogeneous. For SIT, improvements in peripheral regions coexist with increased errors over part of the central ice pack; its RMSE rises from 0.1712 to 0.1762~m despite the lower MAE. Spatial-gradient absolute errors decrease by 13.6\% for SIC and 11.6\% for SIT, while changing by +0.12\% and +0.56\% for temperature and salinity. Taken together, the results identify refinement as a major contributor to the accuracy of the final gridded fields, especially SIC.

\begin{table}[t]
    \centering
    \caption{Effect of grid refinement. Area-weighted scores are averaged over D1--D7 and ten depths (for T/S). Units: pp, $^\circ$C, $10^{-3}$, and m.}
    \label{tab:exp_refinement}
    \footnotesize
    \setlength{\tabcolsep}{2.4pt}
    \renewcommand{\arraystretch}{1.08}
    \begin{tabular}{lrrrrrrrr}
        \toprule
        & \multicolumn{4}{c}{RMSE $\downarrow$} & \multicolumn{4}{c}{MAE $\downarrow$} \\
        \cmidrule(lr){2-5}\cmidrule(lr){6-9}
        Output & SIC & Temp. & Sal. & SIT & SIC & Temp. & Sal. & SIT \\
        \midrule
        Pre-refine. & \underline{14.541} & \underline{1.126} & \underline{0.575} & \textbf{0.1712} & \underline{8.915} & \underline{0.867} & \underline{0.386} & \underline{0.0961} \\
        Refined & \textbf{8.610} & \textbf{0.887} & \textbf{0.534} & \underline{0.1762} & \textbf{2.816} & \textbf{0.552} & \textbf{0.310} & \textbf{0.0705} \\
        \bottomrule
    \end{tabular}
\end{table}

\subsubsection{Contribution of Staged Training}

\begin{figure}[t]
    \centering
    \includegraphics[page=1,width=0.9\columnwidth]{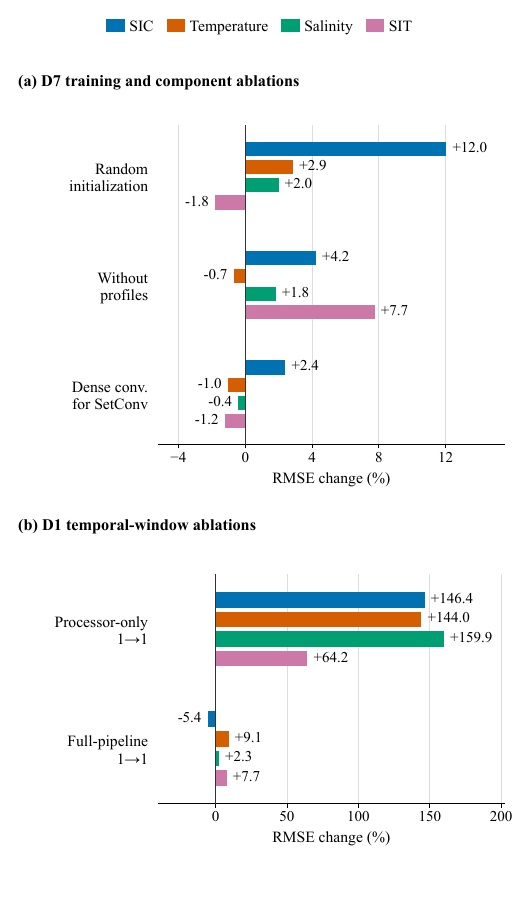}
    \caption{Training and architecture ablations. In reading order, panels compare random initialization, retraining without profiles, dense rasterization in place of SetConv, processor-only $1\rightarrow1$ training, and full-pipeline $1\rightarrow1$ training against their staged reference. The first three comparisons use D7, and the final two use D1. Bars show $100(E_{\mathrm{variant}}/E_{\mathrm{reference}}-1)$; positive values indicate higher RMSE. All results use pixel-weighted evaluation.}
    \label{fig:training_ablation}
\end{figure}

Fig.~\ref{fig:training_ablation} and Table~\ref{tab:exp_training_ablation} show that staged training improves SIC, temperature, and salinity forecasts relative to joint training from random initialization. SIC RMSE decreases from 9.842 to 8.785~pp, a reduction of 10.7\%, while temperature and salinity RMSE decrease by 2.8\% and 2.0\%, respectively. SIT RMSE changes from 0.1614 to 0.1644~m. The largest gain again occurs in SIC, supporting the use of module-wise pretraining before jointly optimizing the observation-to-forecast pipeline.

\subsubsection{Temporal-Window Design}

We compare the standard $7\rightarrow7$ configuration with two $1\rightarrow1$ variants at the common D1 lead. When only the processor is retrained and the encoder and decoder remain frozen, errors increase substantially for all four variables. Retraining the encoder and decoder followed by joint fine-tuning of the full $1\rightarrow1$ pipeline markedly reduces these errors, highlighting the importance of adapting the complete pipeline to the temporal configuration.

The fully trained $1\rightarrow1$ model achieves lower D1 SIC RMSE than the $7\rightarrow7$ model, at 7.569 versus 8.000~pp. However, its temperature, salinity, and SIT RMSE are 9.1\%, 2.3\%, and 7.7\% higher, respectively. The seven-day configuration therefore combines lower D1 errors for three physical-variable groups with direct prediction of the full seven-day horizon.

\subsubsection{Observation Encoding and Profile Inputs}

Replacing SetConv with nearest-grid-cell rasterization and ordinary convolution increases D7 SIC RMSE from 8.785 to 8.995~pp. The replacement uses the same convolutional kernel size, averages observations within each cell, and fills unobserved cells with zero. Temperature, salinity, and SIT RMSE decrease slightly under this replacement, by approximately 1.0\%, 0.4\%, and 1.2\%, respectively. The main benefit of SetConv in this experiment is therefore its contribution to SIC prediction.

The profile-input experiments distinguish the effects of training with profiles from those of removing them after training. Retraining without historical ITP/WOD inputs increases D7 SIC, salinity, and SIT RMSE by 4.2\%, 1.8\%, and 7.7\%, respectively, while temperature RMSE decreases by 0.7\%. In contrast, removing profiles only at inference changes gridded RMSE by less than 0.02\% for all four variable groups under the evaluated input availability. Together, these results show that the benefit associated with profile-enabled training is not equivalent to the immediate sensitivity of the trained gridded predictor to profile removal.

\begin{table}[t]
    \centering
    \caption{Training and architecture ablations (pixel-weighted RMSE; units: pp, $^\circ$C, $10^{-3}$, and m).}
    \label{tab:exp_training_ablation}
    \footnotesize
    \setlength{\tabcolsep}{3.5pt}
    \renewcommand{\arraystretch}{1.08}
    \begin{tabular}{llrrrr}
        \toprule
        Lead & Configuration & SIC & Temp. & Sal. & SIT \\
        \midrule
        D7 & Staged $7\rightarrow7$ & \textbf{8.785} & 0.947 & \underline{0.525} & 0.164 \\
        D7 & Random initialization & 9.842 & 0.974 & 0.535 & \textbf{0.161} \\
        D7 & Without profiles & 9.156 & \underline{0.940} & 0.534 & 0.177 \\
        D7 & Dense conv. (no SetConv) & \underline{8.995} & \textbf{0.937} & \textbf{0.522} & \underline{0.162} \\
        \midrule
        D1 & Staged $7\rightarrow7$ & \underline{8.000} & \textbf{0.918} & \textbf{0.522} & \textbf{0.160} \\
        D1 & $1\rightarrow1$, processor only & 19.714 & 2.241 & 1.357 & 0.262 \\
        D1 & $1\rightarrow1$, full pipeline & \textbf{7.569} & \underline{1.002} & \underline{0.534} & \underline{0.172} \\
        \bottomrule
    \end{tabular}
\end{table}

\subsubsection{Sensitivity to ITP Buoy Availability}

We further examine observation sensitivity by varying the number of retained ITP buoys in four training-period windows initialized on September 15, 2007; October 3, 2008; September 12, 2013; and October 11, 2019. Their maximum available buoy counts are 13, 14, 11, and 11, respectively. Model weights and future evaluation targets remain fixed. Each nonzero count uses three observation draws, while the zero-buoy configuration is evaluated once. Historical inputs in this experiment contain identifiable ITP platforms, with WOD excluded and the profile-token cap fixed at 512.

\begin{figure*}[t]
    \centering
    \includegraphics[page=2,width=0.8\textwidth]{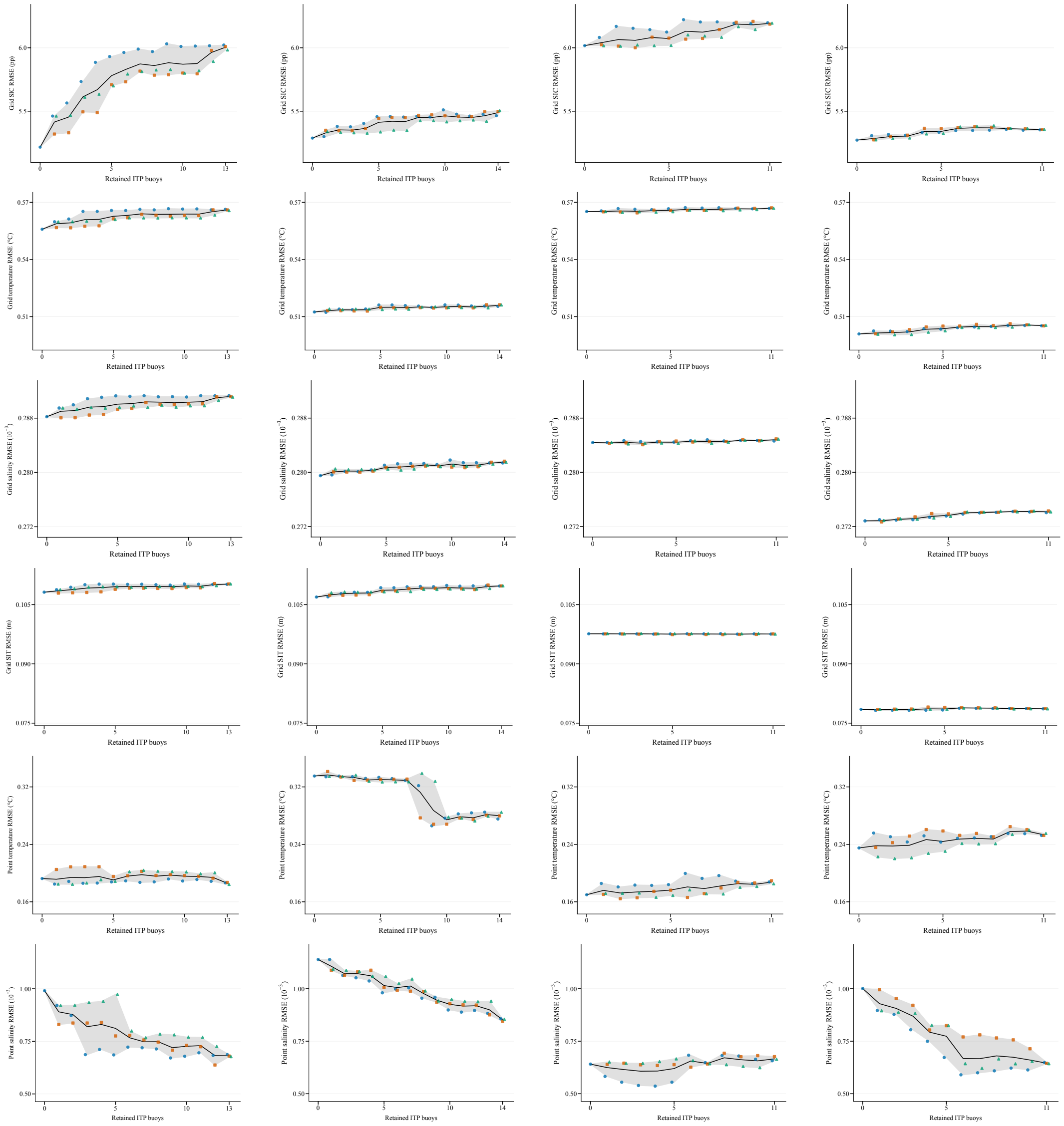}
    \caption{Sensitivity to retained ITP buoys in four windows (columns: Sep~15, 2007; Oct~3, 2008; Sep~12, 2013; Oct~11, 2019). Rows show gridded SIC, temperature, salinity, and SIT RMSE, and pointwise T/S RMSE (colored markers: three draws; black lines: mean; gray bands: min--max).}
    \label{fig:buoy_sensitivity}
\end{figure*}

Fig.~\ref{fig:buoy_sensitivity} shows that the strongest positive responses occur in pointwise salinity. Relative to zero buoys, using the maximum available count reduces salinity RMSE by 31.2\%, 25.2\%, and 35.5\% in the 2007, 2008, and 2019 windows, respectively. The 2013 window exhibits a 3.9\% increase. Pointwise temperature improves by 3.4\% and 16.5\% in the first two windows but deteriorates by 10.3\% and 7.8\% in the latter two.

The gridded and pointwise outputs respond differently to additional profiles. At the maximum buoy count, gridded SIC RMSE increases by 15.1\%, 3.8\%, 2.9\%, and 1.6\% across the four windows. The gains in local salinity prediction therefore do not translate directly into gains in gridded sea-ice prediction. These experiments complement the point-head comparison by showing how local salinity accuracy responds to the availability and spatial composition of profile observations.

Taken together, the ablation results support the integrated design of Arctic E2E for joint sea-ice and upper-ocean forecasting. Grid refinement provides the largest improvements in SIC, temperature, and salinity RMSE while reducing MAE for all four variables. Staged training further improves SIC, temperature, and salinity relative to training from random initialization. Removing profile inputs during training degrades SIC, salinity, and SIT forecasts, whereas replacing SetConv primarily affects SIC. The seven-day formulation also achieves lower day-1 temperature, salinity, and SIT errors than the one-day formulation. These complementary gains support the complete configuration, with particularly consistent benefits for SIC across the refinement, training, and observation-encoding ablations.

\subsection{Seasonal and Extreme-Event Performance}
\label{sec:exp_extremes}

We assess D7 sea-ice forecasts across seasons and rapid ice-change conditions. Windows are grouped by their verification dates, giving 227 DJF, 276 MAM, 276 JJA, and 182 SON cases. Encoder persistence provides the observation-derived reference for forecast evolution, while GLORYS persistence holds the final historical reanalysis field fixed as an additional reanalysis-initialized reference.

\begin{figure}
    \centering
    \includegraphics[width=0.9\columnwidth]{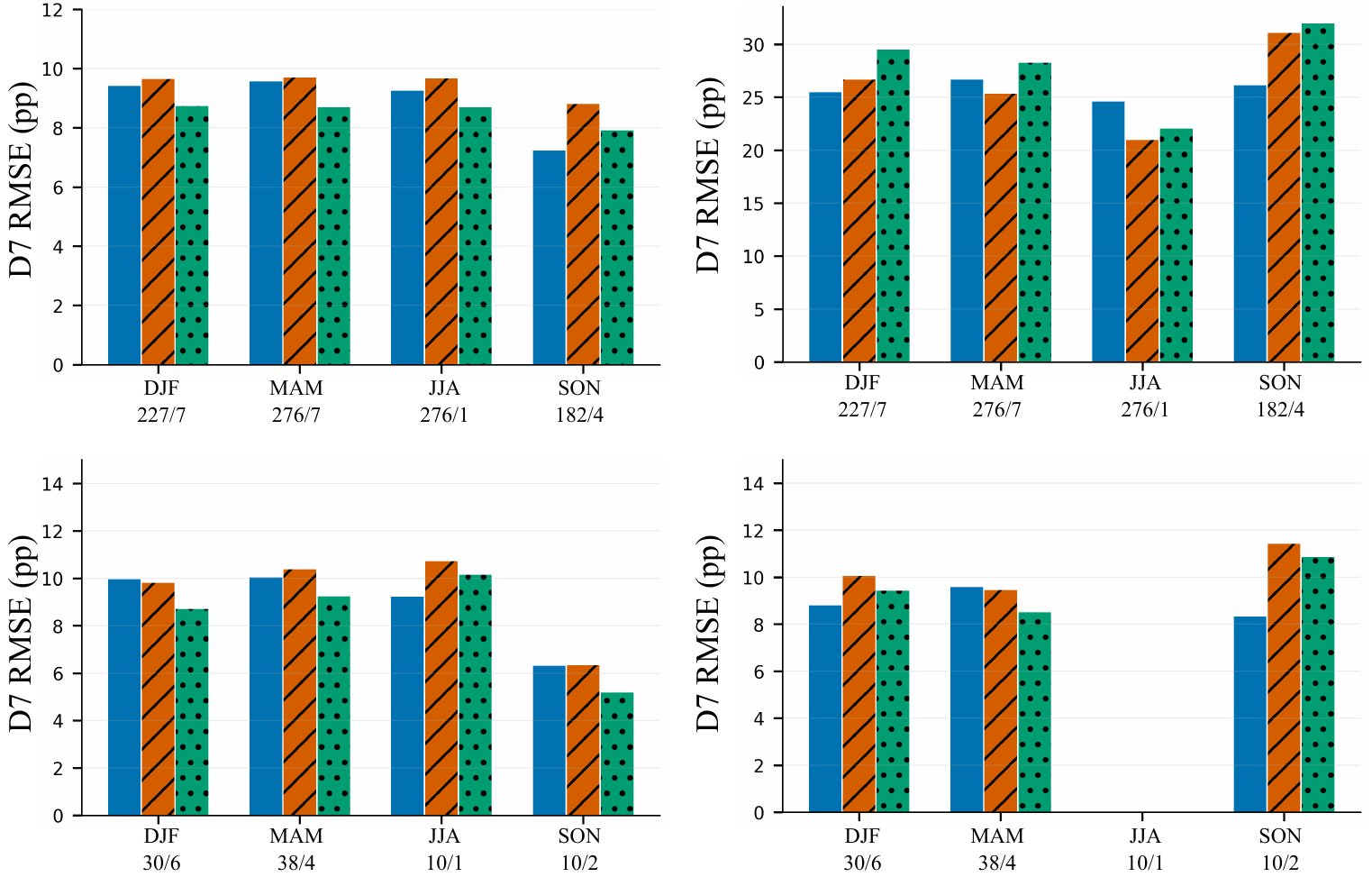}
    \caption{Seasonal and extreme-event D7 SIC RMSE. Panels show full-domain seasonal performance, MIZ performance, extreme retreat, and extreme advance in reading order. Blue bars denote Arctic E2E, orange hatched bars encoder persistence, and green dotted bars GLORYS persistence. Axis annotations report the number of windows and merged extreme-event episodes represented in each group. Episodes are formed by merging overlapping or adjacent intervals from initialization minus six days to initialization plus seven days; these counts do not imply statistical independence. The JJA advance group has no qualifying windows.}
    \label{fig:seasonal_extreme}
\end{figure}

Fig.~\ref{fig:seasonal_extreme} and Table~\ref{tab:exp_extremes} show that Arctic E2E reduces full-domain SIC RMSE relative to encoder persistence in every season. The largest improvement occurs in SON, where RMSE decreases from 8.801 to 7.226~pp, a reduction of 17.9\%. The corresponding reductions in DJF, MAM, and JJA are 2.4\%, 1.4\%, and 4.3\%. Arctic E2E also outperforms GLORYS persistence in SON, indicating a particularly favorable result for autumn sea-ice evolution.

The MIZ results provide a more localized assessment. In DJF and SON, Arctic E2E reduces RMSE relative to encoder persistence from 26.683 to 25.470~pp and from 31.081 to 26.116~pp, respectively. Encoder persistence performs better in MAM and JJA. Relative to GLORYS persistence, Arctic E2E achieves lower MIZ error in DJF, MAM, and SON. Thus, the strongest seasonal gains are accompanied by improvements near the ice edge, particularly in autumn.

Extreme-retreat and extreme-advance windows are defined using seasonal thresholds of seven-day reference sea-ice extent changes estimated from 2006--2022. Retreat requires a negative change below the seasonal 10th percentile, and advance requires a positive change above the 90th percentile. Overlapping or adjacent history-plus-forecast intervals are merged into event episodes, whose counts are reported together with the window counts.

Arctic E2E shows its largest extreme-event gains during winter and autumn ice advance. In DJF, its SIC RMSE is 8.800~pp, compared with 10.052~pp for encoder persistence, a reduction of 12.5\%. In SON, RMSE decreases from 11.417 to 8.326~pp, a reduction of 27.1\%. Arctic E2E also outperforms GLORYS persistence in both groups. In MAM advance, its RMSE is 9.581~pp, compared with 9.450~pp for encoder persistence.

During extreme retreat, Arctic E2E improves upon encoder persistence in MAM, JJA, and SON, but not in DJF. For JJA retreat, its RMSE is 9.229~pp, compared with 10.718~pp for encoder persistence and 10.150~pp for GLORYS persistence; this subgroup comprises ten windows from one merged episode. No JJA advance windows satisfy the event definition in the available validation period.

Ice-edge evaluation further supports the improvement over persistence of the observation-derived state. Across all validation windows, Arctic E2E achieves a D7 IIEE of $0.879\times10^{6}$~km$^2$, 9.1\% lower than encoder persistence at $0.968\times10^{6}$~km$^2$. GLORYS persistence obtains $0.680\times10^{6}$~km$^2$. The mean predicted and reference extents are $9.708\times10^{6}$ and $9.325\times10^{6}$~km$^2$, respectively, giving a positive extent bias of $0.383\times10^{6}$~km$^2$. Together, the seasonal, event-based, and ice-edge results show that the learned forecast evolution improves upon encoder persistence, with the most pronounced SIC gains occurring in autumn and during winter and autumn ice advance.

\begin{table*}[t]
    \centering
    \caption{Full-domain D7 SIC RMSE (pp) by season and extreme-event category. $N$ denotes the number of initialization windows, and $E$ the number of merged extreme-event episodes represented in the group. For all-season rows, $E$ counts the extreme episodes present within that season. Best and second-best RMSE values within each row are bold and underlined, respectively. A dash indicates no qualifying windows.}
    \label{tab:exp_extremes}
    \small
    \setlength{\tabcolsep}{8pt}
    \renewcommand{\arraystretch}{1.12}
    \begin{tabular}{llrrrrr}
        \toprule
        Category & Season & $N$ & $E$ & Arctic E2E & Encoder pers. & GLORYS pers. \\
        \midrule
        All & DJF & 227 & 7 & \underline{9.418} & 9.645 & \textbf{8.733} \\
        All & MAM & 276 & 7 & \underline{9.561} & 9.700 & \textbf{8.697} \\
        All & JJA & 276 & 1 & \underline{9.253} & 9.668 & \textbf{8.693} \\
        All & SON & 182 & 4 & \textbf{7.226} & 8.801 & \underline{7.911} \\
        \midrule
        Retreat & DJF & 30 & 6 & 9.964 & \underline{9.801} & \textbf{8.710} \\
        Retreat & MAM & 38 & 4 & \underline{10.033} & 10.376 & \textbf{9.247} \\
        Retreat & JJA & 10 & 1 & \textbf{9.229} & 10.718 & \underline{10.150} \\
        Retreat & SON & 10 & 2 & \underline{6.308} & 6.334 & \textbf{5.171} \\
        \midrule
        Advance & DJF & 29 & 2 & \textbf{8.800} & 10.052 & \underline{9.419} \\
        Advance & MAM & 35 & 3 & 9.581 & \underline{9.450} & \textbf{8.509} \\
        Advance & JJA & 0 & 0 & -- & -- & -- \\
        Advance & SON & 19 & 2 & \textbf{8.326} & 11.417 & \underline{10.862} \\
        \bottomrule
    \end{tabular}
\end{table*}

\section{Discussion}
\label{sec:discussion}
\subsection{Observation-Driven Joint Forecasting}
The results demonstrate that heterogeneous observations can support joint forecasts of Arctic sea ice and the upper ocean without historical reanalysis fields as inference inputs. Arctic E2E achieves its strongest improvements in SIC and thickness while also obtaining the lowest salinity errors among the compared models. These gains indicate that a unified observation-to-forecast framework can accommodate both spatially extensive satellite measurements and irregular profile observations. The distinction between inference inputs and training supervision remains important: the model uses reanalysis fields as gridded training targets, and its grid-based evaluation measures agreement with those fields. Evaluation against in situ observations therefore provides complementary evidence of its ability to predict conditions at sampled locations.

The benefits are not uniform across variables. Although Arctic E2E leads six of the eight main comparison metrics, ConvNeXt achieves lower temperature errors. This contrast shows that jointly forecasting multiple physical variables does not imply identical improvements for every target. It also motivates further investigation of variable-specific representation and optimization within the shared framework. Moreover, joint prediction alone does not establish physical consistency between sea ice, temperature, and salinity. Assessing cross-variable relationships and introducing appropriate physical constraints would help determine whether improved predictive accuracy is accompanied by more consistent sea-ice--ocean evolution.

\subsection{Complementary Contributions of the Model Components}
The ablation results indicate that the complete configuration benefits from components with complementary roles. Grid refinement produces the largest reductions in SIC, temperature, and salinity RMSE, whereas staged training improves these variables relative to training from random initialization. Profile inputs contribute to SIC, salinity, and thickness accuracy in the retraining experiments, while SetConv primarily benefits SIC. The seven-day formulation additionally improves day-1 temperature, salinity, and thickness forecasts relative to the one-day formulation. Together, these findings support the integrated design without requiring every component to improve every variable.

The difference between MAE and RMSE responses provides further insight into the refinement head. For SIT, refinement reduces MAE by 26.7\% but increases RMSE by 2.9\%. Because RMSE places greater weight on large errors, this combination is consistent with lower average absolute error accompanied by a less favorable distribution of large residuals. The spatial results, which show degraded thickness predictions in parts of the central ice-covered region, reinforce the need to assess both aggregate accuracy and regional error structure. Refinement is therefore most clearly beneficial for SIC, temperature, and salinity, while thickness prediction would benefit from additional attention to large local errors.

\subsection{Sparse Observations and Pointwise Prediction}
The coordinate-conditioned decoder provides a distinct advantage for salinity at observation locations, reducing RMSE by 23.2\% relative to interpolation of the predicted grid fields on the same matched samples. This result shows that pointwise decoding can provide information not fully expressed by the gridded output and its subsequent interpolation. However, the corresponding temperature RMSE increases by 5.3\%, indicating that the value of direct pointwise prediction depends on the target variable. The available observations establish this benefit for salinity at sampled locations rather than uniformly throughout the Arctic water column.

The profile-input experiments also distinguish the effects of training with observations from sensitivity to their removal during inference. Retraining without profiles degrades several gridded forecast variables, whereas suppressing profiles only during inference changes the aggregate gridded errors very little. These experiments measure different effects and do not, by themselves, identify the mechanism responsible for the difference. The buoy-sampling experiments further show that additional profiles can substantially improve pointwise salinity while producing mixed responses in other outputs. Since changing buoy availability also changes spatial coverage and the observations retained under the token limit, these results highlight the importance of observation placement and selection alongside observation count.

\subsection{Seasonal Dependence and Future Directions}
The seasonal results suggest that the benefits of learned evolution depend on the forecast regime. Arctic E2E improves day-7 SIC RMSE over observation-encoder persistence in every season, with the largest reduction of 17.9\% in autumn. Improvements during rapid ice-advance conditions in winter and autumn further identify growing-ice conditions as a promising application of the framework. Nevertheless, performance in the MIZ is less consistent than domain-wide performance, emphasizing that accurate large-scale concentration fields do not necessarily imply equally accurate ice-edge placement.

This distinction is also reflected in the ice-edge diagnostics. Arctic E2E reduces day-7 IIEE by 9.1\% relative to observation-encoder persistence, but its mean predicted ice extent remains higher than the reference. Future development should therefore address ice-edge location and extent bias alongside grid-cell errors. Evaluation on independent years not used for model selection, broader profile coverage, and additional extreme events will be important for assessing generalization. Longer forecast horizons and targeted physical constraints offer further opportunities to extend the framework while preserving the advantages demonstrated for seven-day observation-driven prediction.

\section{Conclusion}
\label{sec:conclusion}
This study presents Arctic E2E, an end-to-end framework that directly forecasts Arctic sea-ice and upper-ocean conditions from heterogeneous satellite and in situ observations. By integrating observation encoding, spatiotemporal prediction, and grid- and point-based decoding, the framework jointly predicts SIC, SIT, potential temperature, and salinity over a seven-day horizon without requiring historical reanalysis fields as forecast inputs. This formulation connects irregularly sampled observations with spatially continuous, multivariable forecasts within a unified model.

Experiments demonstrate the effectiveness of this observation-driven approach. Arctic E2E achieves the lowest errors in six of the eight RMSE and MAE comparisons across the four target variables. Relative to the best competing baseline for each metric, it reduces SIC and thickness RMSE by 32.0\% and 22.3\%, respectively, while also achieving the lowest salinity RMSE and MAE. The coordinate-conditioned point decoder further reduces salinity RMSE by 23.2\% compared with interpolation of the predicted grid fields. Ablation studies support the complementary contributions of grid refinement, staged training, and observation encoding, with particularly consistent benefits for SIC. Seasonal evaluations additionally show lower day-7 SIC RMSE than observation-encoder persistence in all four seasons, with the largest improvement in autumn.

Overall, these findings establish the potential of directly integrating heterogeneous observations for joint Arctic sea-ice and upper-ocean forecasting. Future work will focus on improving temperature prediction and ice-edge evolution, strengthening the use of sparse profile observations, and evaluating generalization to independent years and longer forecast horizons.


\bibliography{references}
\bibliographystyle{IEEEtran}

\vfill

\end{document}